\documentstyle[twoside,fleqn,espcrc2,psfig,here]{article}
\newcommand{\be}{\begin{eqnarray}}
\newcommand{\ee}{\end{eqnarray}}
\newcommand{\ba}{\begin{array}}
\newcommand{\ea}{\end{array}}
\newcommand{\la}{\langle}
\newcommand{\ra}{\rangle}
\newcommand{\no}{\nonumber}
\newcommand{\Op}{{\cal O}}
\title{$B^0$-$\bar{B}^0$ mixing with quenched lattice
       NRQCD\thanks{Presented by N.Yamada.}
      }
\author{
  JLQCD Collaboration:
  N.~Yamada\address{High Energy Accelerator Research Organization~(KEK),
                       Tsukuba, Ibaraki 305-0801, Japan},
  S.~Aoki\address{Institute of Physics, University of Tsukuba, 
                  Tsukuba, Ibaraki 305-8571, Japan},
  R.~Burkhalter$^{\rm b,}$\address{Center for Computational Physics, University of Tsukuba,
                        Tsukuba, Ibaraki 305-8571, Japan},
  M.~Fukugita\address{Institute for Cosmic Ray Research, 
                      University of Tokyo, Kashiwa, 277-8582, Japan},
  S.~Hashimoto$^{\rm a}$,
  K-I.~Ishikawa$^{\rm a}$,
  N.~Ishizuka$^{\rm b,c}$,
  Y.~Iwasaki$^{\rm b,c}$,
  K.~Kanaya$^{\rm b,c}$,
  T.~Kaneko$^{\rm a}$,
  Y.~Kuramashi$^{\rm a}$,
  M.~Okawa$^{\rm a}$,
  T.~Onogi\address{Department of Physics, Hiroshima University,
                   Higashi-Hiroshima, 739-8526, Japan},
  S.~Tominaga$^{\rm c}$,
  N.~Tsutsui$^{\rm a}$,
  A.~Ukawa$^{\rm b,c}$,
  T.~Yoshi\'e$^{\rm b,c}$
  }
\begin{document}
\begin{abstract}
  We present our recent results for the $B$-parameters, which parameterize the $\Delta B$=2
  transition amplitudes. Calculations are made in quenched QCD at $\beta$=5.7,
  5.9, and 6.1, using NRQCD for heavy quark and the $O(a)$-improved action
  for light quark. The operators are perturbatively renormalized including
  corrections of $O(\alpha_s/am_Q)$. We examine scaling behavior of the
  $B$-parameters in detail, and discuss the systematic uncertainties using
  scatter of results with different analysis procedures adopted. As a result,
  we find $B_{B_d}(m_b)=0.84(2)(8)$, $B_{B_s}/B_{B_d}=1.017(10)(^{+4}_{-0})$
  and $B_{S_s}(m_b)=0.87(1)(9)(^{+1}_{-0})$ in the quenched approximation.
\vspace{-120mm}
\begin{flushright}
\large KEK-CP-104
\end{flushright}
\vspace{108mm}
\end{abstract}
\maketitle
\section{Introduction}

The precise determination of $\Delta B$=2 transition amplitudes
would allow us to explore new physics beyond the Standard Model.
A number of works to compute the amplitudes have been already done
and are still in progress using the lattice QCD~\cite{shoji}.
The $\Delta B$=2 transition amplitudes are conventionally expressed
in terms of the corresponding $B$-parameters,
which parameterize the deviation from the vacuum saturation approximation
of the amplitudes.
Here we present our recent results on the two phenomenologically important
$B$-parameters, $B_{B_q}$ and $B_{S_q}$,
which are defined in the $\Delta B$=2 transitions
with $\Op_{L}=\bar{b}\gamma_\mu P_L q\ \bar{b}\gamma_\mu P_L q$ and
$\Op_{S}=\bar{b} P_L q\ \bar{b} P_L q$, respectively,
where $q$=$s$ or $d$ and $P_{L}=1-\gamma_5$.

One of the features in this work is the use of lattice NRQCD~\cite{nrqcd} for heavy quark,
which enables us to obtain a better control of a systematic error than
other approaches since it does not need any extrapolation in heavy quark mass.
The price one has to pay, however, is that in using effective theory
one can not take the continuum limit and so that
the power-law divergence of the form $\alpha_s^n/(am_b)^l$ and
the discretization error of $a^2\Lambda_{\rm QCD}^2$ appear simultaneously.
This means that we have to choose lattice spacing such that
both sources of systematic errors are under control.
To see this we perform quenched QCD simulations at $\beta$=5.7, 5.9, and 6.1 and
investigate stability of the results against the lattice spacing.

The simulation parameters are summarized in Table~\ref{tab:sim_para}.
We use the standard plaquette action for gauge field and the $O(a)$-improved action
for light quark.
In order to see the chiral behavior,
we take four different masses for light quarks in the range $[m_s/2,m_s]$.
We also take five different masses for heavy quarks in the range $[2m_b/3,4m_b]$
to investigate the convergence of $1/m_Q$ expansion.
The lattice spacing for each lattice, shown in Table~\ref{tab:sim_para}, is
determined by the string tension.
\begin{table}
  \begin{tabular}{c|ccc}
    $\beta$    & 6.1     & 5.9     & 5.7 \\ \hline
    size       & $24^{3}\times 64$ & $16^{3}\times 48$ & 
                 $12^{3}\times 32$ \\ \hline
    \# of conf.& 518     & 419     & 420     \\ \hline
    $1/a_\sigma$ [GeV]    & 2.29    & 1.64    & 1.08
  \end{tabular}
  \caption{Lattice parameters in this work.}
  \label{tab:sim_para}
\vspace{-5mm}
\end{table}

\section{Method}

The operators $\Op_X$ in the continuum are written in terms of those
on the lattice $\Op_X^{\rm lat}$ as
\be
  \Op_X
= \Op_X^{\rm lat} + \sum_Y z_{X,Y} \Op_Y^{\rm lat}.
\label{eq:op_mat}
\ee
The one-loop matching coefficients $z_{X,Y}$ ($X$=$L$, $S$)
as well as the precise definitions of operators
and other details can be found in Ref.~\cite{pert_mat_nrqcd}, where
the perturbative calculations are performed including $O(\alpha_s/am_Q)$
corrections.

The $b$ quark field defining each operator such as $\Op_L$ and $\Op_S$
is related to the non-relativistic field $h$=$(Q\ \chi^\dag)^{\rm t}$
through the inverse of Foldy-Wouthuysen-Tani transformation as
\be
b = \left[ 1 - \frac{\vec{\gamma}\cdot\vec{D}}{2m_Q}\right]h.
\ee
It should be noted that
because of a lack of the matching coefficients at the higher order of
$O(\alpha_s\Lambda_{\rm QCD}/m_Q)$ and $O(\alpha_s a\Lambda_{\rm QCD})$,
replacing ${\Op}_Y^{\rm lat}$ in eq.~(\ref{eq:op_mat}) by
an alternative set of operators
${\Op'}_Y^{\rm lat}=\bar{h}\Gamma_Y q\ \bar{h}\Gamma_Y q$
is equivalent to eq.~(\ref{eq:op_mat}) itself up to the higher order corrections.

In the following, we explain our method by taking $B_B$ as an example.
First we define the ``lattice $B$-parameter'' by
\be
  B_Y^{\rm lat}
= \frac{\la\bar{B}^0|\Op^{\rm lat}_Y|B^0\ra}
       {\frac{8}{3}
        \la\bar{B}^0| A_0^{\rm lat}|0\ra
        \la 0| A_0^{\rm lat}|B^0\ra},
\label{eq:def_lat_B_1}
\ee
where $A_0^{\rm lat}$=$\bar{b}\gamma_0\gamma_5 q$.
${B_Y'}^{\rm lat}$ is also defined in a similar manner but with ${\Op'}^{\rm lat}_Y$
and ${A'}_0^{\rm lat}$=$\bar{h}\gamma_0\gamma_5 q$.
The $B_B$ is, then, calculated in the following four forms,
\begin{description}
\item 1: $B_B = \left(B_L^{\rm lat} + \sum_Y z_{L,Y} B_Y^{\rm lat}\right)/(1+z_A)^2$
\item 2: $B_B = B_L^{\rm lat} + \sum_Y z_{L,Y-A^2} B_Y^{\rm lat}$
\item 3: $B_B = \left( B_L^{\rm lat} + \sum_Y z_{L,Y} {B'}_Y^{\rm lat} \right)/(1+z_A)^2$
\item 4: $B_B = B_L^{\rm lat} + \sum_Y z_{L,Y-A^2} {B'}_Y^{\rm lat}$,
\end{description}
where $z_A$ is the one-loop matching coefficient of axial vector current and
$z_{L,Y-A^2}=z_{L,Y}$ except for $z_{L,L-A^2}=z_{L,L}-2z_A$.
The difference between the method 1 and 2 (or between the method 3 and 4)
is of $O(\alpha_s^2)$, since in the former expression the factor $1/(1+z_A)^2$,
which arises as the renormalization factor of $A_0^{\rm lat}$ in the denominator,
is kept untouched and in the latter it is expanded to leave only the linear term in $\alpha_s$.
On the other hand, the difference between the method 1 and 3 (or between the method 2 and 4)
is of $O(\alpha_s p/m_Q)$ or $O(\alpha_s^2/am_Q)$, since the former uses $\Op_Y^{\rm lat}$ and
the latter ${\Op'}_Y^{\rm lat}$.
Note that all these definitions of $B_B$ have equal accuracy up to higher order
contributions as just discussed.
A typical size of these uncertainties may be estimated from scatter of the results.
For the coupling constant, which is included in $z_{X,Y}$,
$\alpha_V(q^*)$ with $q^*=2/a$ is used throughout this work.

In the case of $B_S$, the constant $8/3$ and axial vector current $A_0^{\rm lat}$ in
the denominator of eq.~(\ref{eq:def_lat_B_1}), $z_A$ and $z_{L,Y-A^2}$
are replaced by $5/3$ and pseudo-scalar current $P^{\rm lat}=\bar{b}\gamma_5 q$,
$z_P$ and $z_{S,Y-P^2}$, respectively.

\section{Results}

\begin{figure}
\leavevmode\psfig{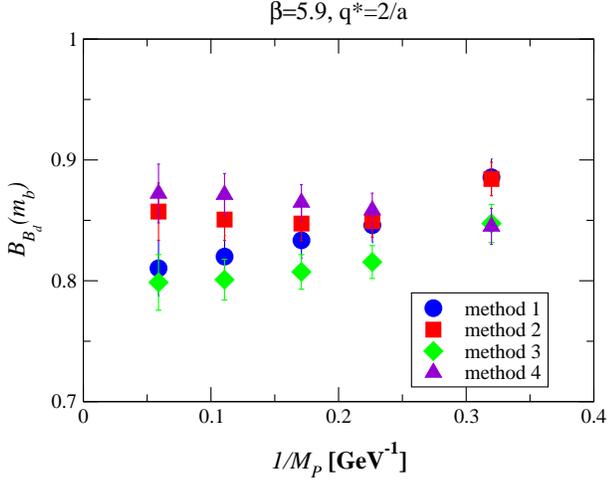}
\vspace{-13mm}
\caption{$1/M_P$ dependence of $B_{B_d}(m_b)$ at $\beta$ = 5.9.}
\label{fig:BBd_5.9}
\vspace{-5mm}
\end{figure}

\begin{figure}[t]
\leavevmode\psfig{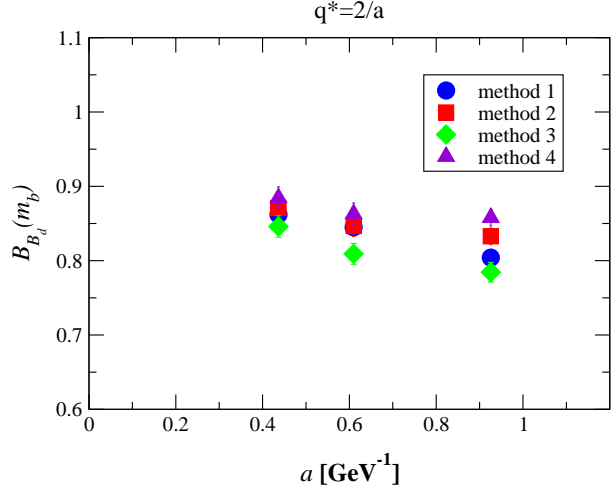}
\vspace{-13mm}
\caption{$B_{B_d}(m_b)$ as a function of lattice spacing.}
\label{fig:a_dep_BBd}
\vspace{-5mm}
\end{figure}

In the following, we show the results of
$B_{B_d}(m_b)$, $B_{B_s}/B_{B_d}$ and $B_{S_s}(m_b)$ obtained with the four methods.
Figure \ref{fig:BBd_5.9} shows the $1/M_P$ dependence of $B_{B_d}(m_b)$ obtained at
$\beta$=5.9, where $M_P$ is the pseudo-scalar heavy-light meson mass.
The renormalization scale of the continuum operator is set to $\mu$=$m_b$=4.8 GeV.
First we find that four plots in Fig.~\ref{fig:BBd_5.9} show various $1/M_P$ dependences
and there is no clear tendency.
This means that $B_B$ depends on $1/M_P$ only weakly, at least,
in the region of $m_Q\ge 2m_b/3$.
Next one can see that
the difference between the method 1 and 2 (or 3 and 4) becomes larger
as one goes to the static limit and dominate the scatter of the results near the static limit.
This might be interpreted as a typical size of the $O(\alpha_s^2)$ uncertainty
as discussed before.
We also find a deviation at the lightest $M_P$,
which is dominated by the difference between the method 1 and 3 (or method 2 and 4).
At this lightest $M_P$, the bare heavy quark mass in lattice unit is not much larger than one
($am_Q$=1.3), so we might infer that this difference indicates a typical size of the
$O(\alpha_s^2/am_Q)$ error rather than the $O(\alpha_s p/m_Q)$.
Interpolating all the data to the physical $B$ meson mass,
typical sizes of the uncertainties are estimated as
\be
\left.\ba{c}
O(\alpha_s^2)\sim 5\%  \\
O(\alpha_s^2/am_Q)+O(\alpha_s p/m_Q)\sim 4\%
\ea\right.
\label{eq:scat_sys_err}.
\ee
Repeating the same analysis at three lattice spacings, we obtain the lattice
spacing dependence of $B_{B_d}$, which is plotted in Fig.~\ref{fig:a_dep_BBd}.
From this figure we can see that the three data are consistent with each other within
the magnitude of scatter of the results.
This suggests that the scaling violation for $B_{B_d}$ is not too large.
We take our best estimate from the central value of the results at $\beta$=5.9.
The remaining uncertainty is estimated by the naive order counting, which gives
\be
\hspace*{-7mm}
&& O(\alpha_s^2)     \sim7\%,~
   O(\alpha_s p/m_b) \sim2\%,\no\\
\hspace*{-7mm}
&& O(\alpha_s^2/am_b)\sim4\%,~
   O(a^2p^2)         \sim3\%,~
   O(\alpha_sap)     \sim5\%.\no
\ee
Adding them in quadrature we obtain 10\% as our estimate for the systematic error,
which is more conservative than eq.~(\ref{eq:scat_sys_err}), that is,
just taking scatter of the results as systematic uncertainty.
Our preliminary result in the quenched approximation is $B_{B_d}(m_b)$=0.84(2)(8),
where the errors are the statistical and systematic ones, respectively.

\begin{figure}
\leavevmode\psfig{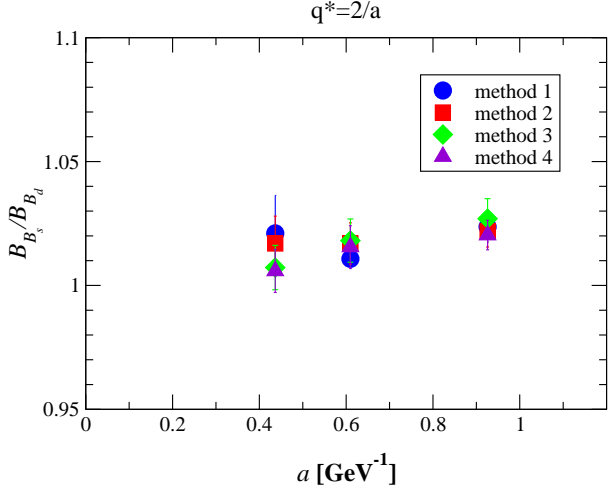}
\vspace{-13mm}
\caption{$B_{B_s}/B_{B_d}$ as a function of lattice spacing.}
\label{fig:a_dep_BsBd}
\vspace{-5mm}
\end{figure}

Repeating the similar analysis, we obtain the lattice spacing dependences of
$B_{B_s}/B_{B_d}$ shown in Fig.~\ref{fig:a_dep_BsBd}.
For this quantity, we observe neither method dependence nor large scaling violation.
We take our best estimate from $\beta$=5.9 lattice as before, and obtain
$B_{B_s}/B_{B_d}=1.017(10)(^{+4}_{-0})$
by only considering the uncertainty in $\kappa_s$.

\begin{figure}[t]
\leavevmode\psfig{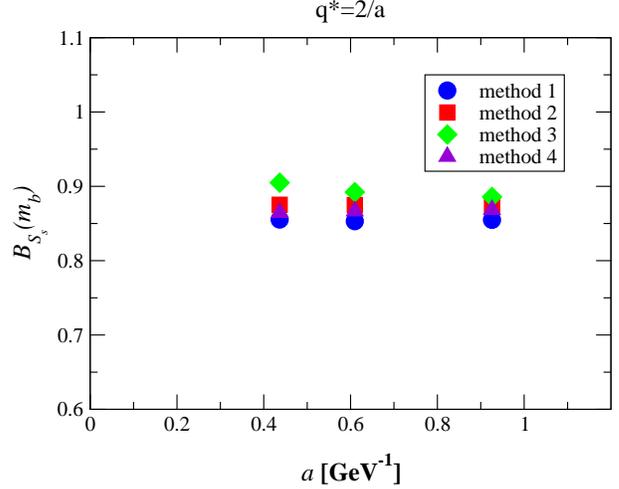}
\vspace{-13mm}
\caption{$B_{S_s}(m_b)$ as a function of lattice spacing.}
\label{fig:a_dep_BSs}
\vspace{-5mm}
\end{figure}

Figure~\ref{fig:a_dep_BSs} shows the lattice spacing dependence of $B_{S_s}$.
The data plotted in this figure does not show a large scaling violation within
the systematic uncertainty indicated by the scatter of the results.
Following the same analysis as in the case of $B_{B_d}$, we obtain
$B_{S_s}$=0.87(1)(9)($^{+1}_{-0}$) in the quenched approximation,
where the errors are statistical, systematic and coming from the uncertainty of $\kappa_s$, respectively.

\section{Applications}

Let us now consider a few applications of our results.
The first one is the mass difference in the neutral $B_q$ meson systems,
which is given by the CKM matrix elements ($|V^*_{tb}V_{tq}|$) and $f_{B_q}^2B_{B_q}$
up to a known factor as
\be
    \Delta M_{B_q}
&\propto&
    |V_{tb}^*V_{tq}|^2
    f_{B_q}^2B_{B_q}.
\label{eq:mass_diff_formula}
\ee
Assuming the additional 10\% uncertainty due~to
the quenched approximation to above result for
$B_{B_d}$ and~the~recent world average of the decay
constant~from~$n_f$=2~simulations~$f_{B_d}$=210(30) MeV,
we obtain $|V^*_{tb}V_{td}|$=0.0074(12).
This result can be used to estimate the frequency of the $B_s^0$-$\bar{B}_s^0$ mixing.
If we take $|V_{ts}|$=$|V_{cb}|$=0.0404(18) and $f_{B_s}/f_{B_d}$=1.16(4)
and assume the uncertainty due to the quenched approximation for $B_{B_s}/B_{B_d}$
to be 10\% again, we obtain $\Delta M_{B_s}$= 20.0(4.7)~ps$^{-1}$.
The current experimental bound is
$\Delta M_{B_s}$ $>$ 14.9~ps$^{-1}$~\cite{ichep00}.

The second application is the width difference of $B_s$ mesons.
The width difference is theoretically calculated using the $1/m_b$ expansion.
Beneke {\it et al.} have completed the formula with the next-to-leading order
correction in Ref.~\cite{beneke}.
Following their analysis but with $f_{B_s}$=245(30)MeV~\cite{shoji},
the formula is given by
\be
  \left(\frac{\Delta\Gamma}{\Gamma}\right)_{B_s}
= \left(\frac{f_{B_s}}{245(30){\rm MeV}}\right)^2
  \Big[\  0.008 B_{B_s}(m_b)                      \no\\
\hspace*{-6mm}
          +\ 0.227(17) B_{S_s}(m_b)
          -  0.086(17)\
    \Big].
\label{eq:width_diff_formula}
\ee
Using the results for $B$-parameters but with the additional 10\% error
due to the quenched approximation, we obtain $(\Delta\Gamma/\Gamma)_s$ = 0.119(29)(32)(17),
where errors come from the uncertainties of $f_{B_s}$, $B_{S_s}$ and the last term
in eq.~(\ref{eq:width_diff_formula}), which shows $O(1/m_b)$ contribution, respectively.
The current experimental bound is
$(\Delta\Gamma/\Gamma)_s$ = 0.16($^{+16}_{-13}$)~\cite{ichep00}.
It would be quite interesting to compare our result with
the coming Tevatron Run II data, which are expected to improve
the current experimental results significantly soon.

\section*{Acknowledgment}
This work is supported by the Supercomputer Project No.54 (FY2000)
of High Energy Accelerator Research Organization (KEK),
and also in part by the Grant-in-Aid of the Ministry of 
Education (Nos. 10640246, 10640248, 11640250,~11640294,~11740162,~12014202,
12640253, 12640279, 12740133).
K-I.I, T.K and N.Y are supported by the JSPS Research Fellowship.



\begin{thebibliography}{99}
\bibitem{shoji}
For reviews, see, for example,
  S. Hashimoto, Nucl. Phys. B (Proc.Suppl.) 83 (2000) 3;
  C. Bernard, these proceedings.
\bibitem{nrqcd}
  B.A.~Thacker and G.P.~Lepage, Phys. Rev. D43 (1991) 196;
  G.P.~Lepage, L.~Magnea, C.~Nakhleh, U.~Magnea, and K.~Hornbostel,
  Phys. Rev. D46 (1992) 4052.
\bibitem{pert_mat_nrqcd}
  S.~Hashimoto, K-I.~Ishikawa, T.~Onogi, M. Sakamoto, N.~Tsutsui, and N.~Yamada,
  Phys. Rev. D62 (2000) 114502; these proceedings.
\bibitem{ichep00}
  A. Golutvin, plenary talk given at the XXXth International Conference on High Energy Physics
  July 27 - August 2, 2000, Osaka, Japan.
\bibitem{beneke}
  M.~Beneke, G.~Buchalla, C.~Greub, A.~ Lenz, and U.~Nierste,
  Phys. Lett. B459 (1999) 631.
\end{thebibliography}
\end{document}